\documentclass[a4paper]{article}
\usepackage[T1]{fontenc}
\usepackage{authblk}
\usepackage[english]{babel}
\usepackage{url}
\usepackage{quoting}
\usepackage{geometry}
\usepackage[hang]{footmisc}
\usepackage{stackengine}
\title{\textbf{A minimalist account of agency in physics}}
\date{}
\begin{document}
\author[a]{Ali Barzegar\thanks{barzegar11@gmail.com}}
\author[b]{Emilia Margoni\thanks{emilia.margoni@phd.unipi.it}}
\author[a]{Daniele Oriti \thanks{daniele.oriti@physik.lmu.de}}
\affil[a]{\small Arnold Sommerfeld Center for Theoretical Physics \& Munich Center for Mathematical Philosophy, Ludwig-Maximilians-Universit\"at M\"unchen, Germany, EU}
\affil[b]{\small Università degli Studi di Firenze/Université de Genève/Università di Pisa}

\maketitle
\vspace{-1cm}
\begin{abstract}
\noindent We adopt a top-down approach to agency aimed at developing a minimalist, scalable and naturalized account of it. After providing a general definition, we explore some possible extensions and refinements, domain of applicability, as well as a comparison with other recent accounts of agency, and possible objections to our proposal. With respect to what we classify as strong (such as Tononi's) and weak (such as Rovelli's) characterizations, our notion of agency situates itself in a middle position -- our intent being precisely that of spelling out the advantages of this median account within a variety of contexts, such as the interpretation of quantum mechanics, the debate on the nature of physical laws and bayesianism. \\ 
\end{abstract}

\section{Introduction}
There is a sense in which agency is everywhere to be found: a river gradually eroding the rocks that mark its banks, squid ink being released to seed a predator, a child knocking over a vase while crawling, thus soiling the carpet. Every time systems interact, they display causal relationships that result in reciprocal change. However, the notion of agency -- which has prospered throughout the last few decades in a variety of disciplines and research areas, such as psychology, anthropology, philosophy, and cognitive neuroscience, as well as physics -- is generally tied to a thinner connotation, whereby an agent gets identified with an entity that is able to act in specific ways with respect to its surrounding environment -- the notion of agency marking the exertion of this specific function. 

\noindent While traditionally the ability to act is related to the performance of intentional faculties -- two leading figures of such a characterization of agency being, within contemporary analytic philosophy, Elizabeth Anscombe (1957) and Donald Davidson (1963) -- more recent debates have centred on rebutting such a restricted view. The underlying assumption is that, whenever attention is drawn to entities simpler than human beings, the traditional view proves too demanding, to such an extent that the notion of agency should be detached from that of representational and intentional states (see, e.g., Silberstein \& Chemero, 2011; Hutto \& Myin, 2014). 

\noindent As no agreed-upon definition is available in the literature, speaking about agency means invoking a variety of kinds and contexts. Some adopted notions are mental, shared, collective, relational, and artificial agency\footnote{For a thorough discussion about the various characterizations of agency, see Schlosser (2019).}, where in all these declinations the invoked notion can be more or less conforming to the traditional (intentional) meaning. For instance, while shared agency refers to the exertion of action by two or more individuals and collective agency occurs when two or more individuals act as a group, it is still debatable whether the resulting action should be reduced to the agency of the single involved individuals, or whether this shared or collective faculty constitutes something on top of individual agency, thus opening the question if, within the traditional paradigm, group of individuals should be accorded intentional or mental states. 

\

\noindent All this comes down to the issue of whether there is a way to provide a basic characterization of agency. Recently, Barandiaran et al. (2009) have worked on what they dub “minimal agency”, a notion that does not require the possession of representational or mental states. According to them, agency is defined whenever a system satisfies three conditions, namely that: a) it establishes its own individuality; b) it displays an interactional asymmetry with respect to its surrounding environment; and c) the interaction between the system and the environment is based on certain normative parameters. On this perspective, organisms such as bacteria meet the minimal criterion of agency, thus enlarging the eligible entities with respect to more traditional accounts. 

\noindent Another recent proposal -- which embraces a computer science perspective -- is Biehl, Ikegami \& Polani (2016), which advances an “information theoretic notion of integrated spatiotemporal patterns” as “the basic building block of an agent definition” (2016, p. 1). Their strategy is the following: 1. define spatiotemporal patterns for both living, such as bacteria and animals, and non-living, such as rocks and crystals, persisting systems; 2. establish some criteria, namely perception, action and goal-directedness, to catalogue those spatiotemporal patterns; 3. identify, among the total set of persisting systems, those that meet the agency criterion.  

\noindent Though intriguing, more flexible and broader as compared to traditional accounts, both these proposals still refer to notions such as perception, action, and intentionality mediated through normative parameters. It follows that the associated accounts of agency can hardly be considered a fully naturalized, let alone minimalist, characterization of it.  

\

\noindent The intent of the present paper is to propose such a minimalist, scalable and naturalized characterization of agency, based on a minimal notion of representation, modelling and normativity. From this minimalist basis, supplementary features can be added which correspond to higher-order levels of complexity. This account will be based on the exertion of modelling capacities that a certain system displays with respect to the surrounding environment, where even the notion of environment will be scrutinized. On this view, modelling is the hallmark of agency: the more complex the model, the more complex the agent. By relying on a minimalist characterization of modelling, our account situates itself between the two extremes of depicting an agent as every possible interacting system -- what we dub a \emph{weak} account of agency, such as Rovelli's (2020) -- or of taking only a very narrow set of systems as agents in virtue of the possession of some very specific features -- what we dub a \emph{strong} account of agency, such as Tononi's (2015) and collaborators' (Albantakis et al. 2020) requirement that an agent should display some minimal grade of consciousness.

\

\noindent The paper is structured as follows. In the section 2, we provide our definition of agency and discuss its main features. The 3rd section expands on its domain of applicability, as well as further extension and refinements. In the 4th section, a comparison between our account and other recent proposals is outlined. The 5th section addresses some objections that can be raised against our proposal. Finally, we draw some concluding remarks and provide further lines of inquiry. 

\section{A minimalist definition of agency}

What is it that we are talking about when making reference to agents? Though currently available literature shows a proliferation of very different characterizations, an agent can be generally identified with a system or entity that interacts -- in a way that needs to be further specified -- with the surrounding environment (see, e.g., Ried et al., 2019). Zooming in from this coarse-grained understanding, we need to spell out more thoroughly which subset of systems or entities actually count as agents. That is, if an agent is a system situated within an environment, interacting with and acquiring information about it, the question becomes what is the hallmark of agency \emph{vis-à-vis} generic interactions. Here is our tentative definition:\\

\noindent \emph{An agent is an information-processing system which is characterized by modelling activity with respect to its surrounding environment, geared towards successive interaction.\\}

\noindent Let us emphasize some preliminary aspects of this definition\footnote{For this definition to be relevant for the understanding of scientific epistemology and more generally scientific activity, it should allow an agent to be able to model at least some of its own (scientific) activity (E. Curiel, private communication). In more detail (E. Curiel, 2020), the empirical content of scientific propositions is grounded in an agent's capacity to schematize the observer in the models of experiments, where the notions of observer and experiment should be conceived broadly. However, our point, as will be discussed afterwards, is that, moving from this minimal characterization, a hierarchy of ‘agency level’ can be defined, Curiel's requirements being eventually realized from a certain level onward.}:\\

\begin{enumerate}
\item First of all, note that this definition is a functional one: any system exhibiting the required properties can be qualified as an agent, our account being extremely neutral with respect to the nature of the eligible entities. It thus becomes an empirical question to determine which systems satisfy this definition of agency, and which do not. 

\item  Second, it is a {\it minimal}  definition of agency: an agent is an entity that organizes the information from the environment following the interaction with it -- where this organization is what we identify with the agent's modelling capacity, which in turn is based on a minimal notion of representation. Every other feature of agency can be added to this minimal level as a higher-order level of complexity.
An agent is an information-processing system, and part of this processing activity is modelling. The latter provides a basis for successive interactions with the environment, thus implying a minimal notion of normativity, in the basic sense that the modelling activity informs (it is used in) successive interactions. On this approach, one can envisage a spectrum of agency in terms of the way in which the latter reacts to external inputs – with the two extremes of “simple reflexes to environmental stimuli” (Briegel, 2012, p. 2), on the one hand, and the “projected simulation […] into conceivable future situations” (Briegel, 2012, p. 4), on the other. Importantly, both an extremely complex system not interacting with the surrounding environment and a system interacting but devoid of the appropriate internal structure would not meet our criteria for agency. 

\item Third, our definition of agency is meant to be sufficiently scalable and malleable. This means that it occupies a middle position between paradigms defending a strong notion of agency (such as the one which takes the property of, say, having consciousness to be a necessary condition for speaking about agents) and those paradigms envisaging a thin notion of agency (such as the one according to which every interacting system is an agent). Moving from this minimal, but intermediate standpoint, it is then possible to add features of agency that may correspond to stronger notions, or to remove those that would give us back the thinner ones (although in this case, we would argue, we are not considering proper agency anymore). 

\end{enumerate} 

\noindent After this preliminary analysis, it is necessary to unpack the underlying assumptions of our proposed definition. In particular, it is worth exploring the role that modelling and representation play, as well as the physical requirements necessary for their exertion. 

\subsection{A minimalist account of modelling}

Our account associates the capacity of a certain system to model the environment with agency. A minimal notion of agents as entities capable of modelling the environment can be articulated further in the following way. The agent receives data from its interaction with the environment. \footnote{ Schematically, one could interpret these interactions as measurements by the agent on the environment, thereby acquiring the measurement results as data; however, while any measurement is an interaction, not every interaction counts as a measurement, and we are interested in understanding the specific conditions under which a certain interaction can be taken as providing a measurement result.} 
From this perspective, for an agent to be able to model its environment, it should display the capacity to model these data, thereby providing some structure to them. That is, the agent should be able to build \emph{data models}, which corresponds to the minimal level of modelling activity. Minimality is guaranteed by a categorization scheme of the information gathered from the surrounding environment --  this capacity standing for the fact that an agent is able to give some structure to such information.\footnote{For a thorough discussion on the relationship between data and models of data see, e.g., Leonelli (2019).} 

\noindent What, then, is a data model? Formally, a data model can be defined in set-theoretic \footnote{See, among others, Hegner and Maulucci (1978), Yannakoudakis (2007).}  or category-theoretic terms \footnote{See, e.g., Zhao (2019).} as a set of elements or categories with relations holding between them. Based on the number of elements, relations, and degrees of freedom, one can then define a measure of complexity for different data models. For instance, one can start evaluating the complexity of the data model by counting the number of involved elements. Successively, one can consider binary relations, then binary plus tertiary relations, and so on, thus adding levels of complexity.

\

\noindent Let us turn to a more concrete analysis of data models. Obviously, they are models of \emph{data}. However, it should be noted that the notion of data, and especially \emph{raw data}, is an utterly ambiguous one. 
This is related to the old debate in methodology on the theory-ladenness of observations and empirical facts. \footnote{As a particularly instructive discussion on the topic, see Ian Hacking's (2002; 2009).} Furthermore, there is no one-to-one correspondence between data sets and data models. There are always competing patterns within data, which implies the adoption of criteria to extract data models. At a first level of analysis, what you see in a data model is a minimal level of filtering with respect to raw data or in the patterns among data in which not all the data are included, as some of them may not be useful and/or reliable. At a higher level, agents use data models in a future-oriented manner. For example, the choice of a data model can be articulated in terms of some subtler function within the agent engaged in the modeling activity such as its goal-orientedness. From this perspective, data models become goal-oriented organizations of data, to such an extent that different degrees of goal-orientedness correspond to different degrees of complexity of the involved agent. At more refined levels of organization, they use data models in counterfactual situations, to predict and respond to possible situations occurring in the future. Within these counterfactual scenarios, and in contrast to data models of agents at lower levels of complexity, these data models and successive elaborations are hypothetical instead of actual.   

\

\noindent In light of what has been discussed so far, an interesting way to elucidate the difference between data models of agents characterized by different levels of complexity is by relying on the relational understanding of data models, according to which they intertwine and so are defined in relation to other models the agent constructs and uses (Leonelli, 2016). From this perspective, there is a feedback loop between different models the agent builds and makes use of. And these other models available to the agent influence the nature and complexity of the data models it is able to construct in the first place. As a consequence of this, by comparing the data models themselves one should be able to distinguish the corresponding agents. This is due to the fact that the more developed agent making use of other models, over and above data models, will construct a more complex data model out of the same data set \emph{vis-à-vis} an agent who just can build and use them. In other words, a difference at the level of data models signals a difference at the level of agency. Moreover, by relying on the relational understanding of models, we can also argue that at a specific level of complexity the model allows for Bayesian probabilistic inferences.\footnote{This is a topic that will be further investigated in the following sections.}  

\

\noindent Following Briegel's (2012), we can provisionally define a first taxonomy of agency, roughly corresponding to three levels of complexity by virtue of the associated modeling activity. At the minimal level, modeling is any kind of structuring or categorization of the data received from the interaction with the environment. Then there is a second level of complexity, which is associated to the existence of feedback loops in the internal information-processing activity of the agent, thus among its constructed (data) models. At a higher level, the agent becomes a self-referential system and provides feedback on its own activity, such as the history of data patterns. When a new input is received from the environment, the agent reacts to it also by considering the stored patterns, not just the currently available ones. At this level, a certain pattern is selected based on the weights the agent assigns to it, thus resulting in a specific activity. 
This third level of complexity can be further expanded if the agent can not only store information, but also elaborate evidence-based future models. In this case, the agent can construct counterfactual scenarios and take advantage of them to simulate or predict its successive interaction with the environment. The more complex the hypothetical patterns, the more complex the agent. 

\subsection{Physical requirements for modelling activity}

So far we have focused our attention on the role played by data models in our minimalist account of agency, and in providing a possible taxonomy of agents, based on their ability to process data collected from the environment. Now we need to spell out what physical requirements enable the production of data models in the first place. According to our definition, information-gathering is a necessary condition for a system to count as an agent. However, not every information-storing system counts as an agent. Over and above gathering and storing information, an agent {\it organizes and then uses} this information for modelling and then successively interacting with the environment (where, again, a minimal notion of modelling is here intended as a categorization of the inputs from the environment). It is then necessary to discriminate between simply acquiring information about the environment and modelling it. The latter activity is not measured by the sheer amount of information one acquires, say by relative information (Rovelli, 2016). 

\noindent Consequently, there should be a physical basis to sustain this activity of organization and structure with respect to the acquired information. Since a minimal form of complexity is required, an agent cannot correspond to an elementary system, namely a system without parts. Importantly, when we talk of complexity as a requirement for agency, we do not simply mean that more than one degree of freedom is involved (though this is a necessary condition). We are also referring to the peculiar way in which the modelling activity arises out of the interaction among the involved degrees of freedom. For instance, Albantakis et al. (2020) argue that one leading aspect of an agent is the fact that the latter displays a form of irreducibility with respect to its parts, to such an extent that the integration and relation between its parts constitute a unitary whole, as “all of its subsets constrain and are being constrained by other subsets within the system above a background of external influences” (p. 2). \footnote{For further discussion on this topic see, among others, Maturana and Varela (1980), Hoel, Albantakis and Tononi (2013) and Marshall et al. (2017).} We concur with this perspective, and argue that (in particular, higher-level) agents are something over and above the sum of their component parts, to such an extent that the agent needs something like a central processing system, based on which subsystems are not independent from one another; rather, there is some internal interaction between its parts. 

\

\noindent To summarize, the type of complexity required to define an agent consists of two elements: a number of degrees of freedom and some internal, specific interaction among them. Again, the more structured such internal interaction, the more complex the associated agent. While for Tononi (2012) this internal interaction is what gives rise to integrated information systems (IIS) that are irreducible to their component parts, the \emph{minimal} modelling activity here invoked requires a minimal form of organization or interaction within the system. At higher levels of agency, this discourse translates into the possible irreducibility of higher-order models into lower-level ones, thus calling for a hierarchy of models with higher-level ones being irreducible to the lower-level ones: each and every level of this hierarchy can be regarded as corresponding to a form of agency, characterized by a specific level of complexity in terms of internal structure and modelling activity.   

\noindent To give a concrete example, according to this characterization, an elementary particle cannot display agency because, by definition, it has no parts. However, a system of, say, four electrons can manifest a certain structure which enables it to model the environment.\footnote{Actually, within our account, as will be discussed later on, even a single electron, provided that it can categorize the surrounding environment with respect to its two subsets of degrees of freedom (position and spin), could in principle be taken as an agent.} Then there should be a non-trivial difference between an elementary and a complex system -- a sort of \lq agency threshold\rq -- after which complexity comes in degrees. 

\

\noindent Note that our account is eminently perspectival: an atom would qualify as an agent insofar as it is regarded as a composition of elementary particles, but if we take it as an individual, then it would correspond to a singular system, to which we would not attribute agency. Therefore, our notion of agency is scalable, whereby there is no need to have a notion of a system which is not theory-laden.  
Within this context, perspectivalism boils down to the fact that the attribution of agency depends on the perspective (such as it is provided itself by a model or a theory) according to which systems are classified as elementary or complex. Whatever, according to that perspective, is defined as elementary cannot produce structured information, in that there is no physical structure to support it. Still, being qualified as a complex system does not automatically imply agency: having parts is not sufficient for displaying structured information (with the notion of structured information giving the minimal notion of modelling the environment by the agent). As an instructive example, consider the case of ants. The latter can be regarded as agents, but so does, in the appropriate context, the anthill, thus leading to a scale-dependent account of agency. In other words, depending on what you are looking at, you get a different notion of agent, which is precisely in this sense perspectival.
\noindent Importantly, this perspectival account of agency also means that an agent is always regarded as a sub-system within a larger system: an agent cannot be completely separate from the environment.  

\subsection{Agency and Model Representation}

According to our definition, agency is to a large extent modelling activity. This requires us to discuss in more detail how models are meant to represent. Frigg and Nguyen (2020) argues that an account of model-representation needs to address five topics. First of all, since a model represents a specific part or aspect of the world, namely its \emph{target system}, one must justify in virtue of what a certain model is able to represent it. This implies answering the \emph{Epistemic Representation Problem}, namely to explain why a certain model (or graph, diagram, drawing) represents a certain target system. Second, as models are not the only epistemic devices, a model-representation should be analyzed within the larger category of epistemic representation, so much so that one needs to clarify what is the specificity of (scientific) model representation \emph{vis-à-vis} other kinds of epistemic representation -- this is the \emph{Representational Demarcation Problem}. Third, model-representation should clarify what styles are there and in which way they can be characterized -- the so-called \emph{Problem of Style}. Fourth, it is necessary to define some \emph{Standards of Accuracy}, namely when it is possible to state that a certain model or epistemic device constitutes an accurate representation of a specific target system under investigation. Finally, a model-representation should address the \emph{Problem of Ontology}, namely to clarify what type of object models are. 

\

\noindent Obviously, the debate on the nature of representation, and the way in which models are connected to representation is extremely vast. Indeed, there are very different ways models are said to represent.\footnote{For an introductory discussion on the topics of models and the relationship they bear with representation see Frigg and Hartmann (2020), Frigg and Nguyen (2021). Other interesting contributions are Morrison (1999), Isaac (2013), Levy (2015), Massimi (2018).} In the present context, we are interested in detailing the relationship between modelling activity and agency. To cut a long story short, here are some of the most accredited characterizations of how models are assumed to represent. The first is that models represent by way of similarity between the model and the target system (see, e.g., Cartwright, 1983; Giere, 1988,  2010; Weisberg, 2012). The second draws from a structural relation of isomorphism between the model and the target system (see, among others, Bokulich (2011), and King (2016) for a critical discussion). But as similarity and isomorphism are symmetric properties – whereas representation is founded on directionality – there is an issue connected to the multiple realizability as typically associated with these features. Anything can be similar to anything else in different ways. As different descriptions give different structures, an underdetermination of the structure from the target system is at stake. The third view of representation is based on the idea that the model should allow for inferences to be drawn about the target system (see, e.g., Verreault-Julien, 2021). However, for these inferences to be useful there should be a link between them and the target system. That is, the representation relation between the model and its target system should allow for inferences which are about future actions. Otherwise, one is \lq trapped inside the model\rq, and cannot ensure that the conclusions inferred from the model are applicable to the target system. If the agent derives conclusions about the target system just at the level of propositions, this would imply the presence of a closed system, and then the only task of the agent becomes to check the coherence of this set of propositions. One can indeed conceptualize a model as a set of propositions linked by logical connectors and then try to check its coherence. 

\

\noindent Keeping this in the background, it is important to outline that the only way to break out of the model is via some empirical connection between it and its target. Therefore, action stands for the representational link between the model and the target system. It is only when we take into consideration the action of the agent building and using the model that we go beyond any kind of abstract inference. Thus, we have to invoke the agent to be able to say that a model represents a target system. We believe this is the way to go to bring in the directionality of model representation. On the one hand, the agent builds and uses the model to successively interact with the environment. On the other hand, to the extent that the model helps the agent to successively interact with the environment, it can be said to provide a directed and/or purposeful representation of the environment.
This is an important point to emphasize: agency requires models to be defined; they are part of its very definition. On the other hand, agency feeds back on the notion of models. However, this does not lead to any circularity in our argument, because this feedback takes place at the level of the representational {\it function} of models, rather than at the level of their definition: it determines their representational capacities.  

\section{Further refinements and possible applications} 

To summarize what has been discussed so far, our account is meant to capture the bare minimum notion of agency, which is in turn based on the bare minimum of modelling activity, such as a simple categorization of external data. From this perspective, any system with more than one degree of freedom can engage, in principle, in modelling activity, thus counting as an agent, provided the remaining ingredients in our definition (such as its use of the constructed model to inform successive interactions with the environment) are also present. For example, an electron with spin and position can categorize its interactions with the surrounding environment based on these two degrees of freedom, and would therefore count as an agent. Two electrons likewise count as an agent. But, as there is something counterintuitive in regarding two electrons -- and even more so a single electron -- as agents, perhaps some explication is required. 

\

\noindent First of all, it is important to clarify whether we need to introduce a further classification on top of the proposed definition of agency, or even make the same definition more restrictive. In the first case, one should simply offer a further classification of different levels of agency, based on different levels of (complexity of) modeling activity. In the second case, we should raise the minimal threshold for agency, under which this notion cannot be applied fruitfully, above what we have identified already. We have explained our minimal notion already excludes ({\it contra} some views in the literature according to which any interacting physical system can be qualified as an agent) a number of physical systems to be considered as agents . As we have argued, these minimal ingredients concern the presence of more than one degree of freedom, internal structure for acquired information, as well as some correlation between previous information and successive evolution. What else can be required?

\noindent An interesting (and already mentioned) reference is Barandiaran et al's (2009), according to which agency should be taken as an “autonomous organization that adaptively regulates its coupling with its environment and contributes to sustaining itself as a consequence”(p. 367). They base their definition on the following necessary conditions:

\begin{enumerate}
\item A system must define its own \emph{individuality}
\item It must be the active source of activity within its environment (\emph{interactional asymmetry})
\item It must regulate this activity in relation to certain norms (\emph{normativity})
\end{enumerate}

\noindent This definition is very close to ours. With respect to these conditions, we have seen that our account makes reference to a minimalist notion of normativity, fixed by the fact that the stored information is then used by the agent to successively interact with the environment. However, their concept of individuality is closely tied to the notion of autonomy -- where an autonomous system is one whose components or parts form an irreducible, interrelated network. 
By properly tuning their analysis to the purpose of our account, a three-level distinction comprising all physical systems can be defined. The hierarchy would then comprise 1. systems 2. autonomous systems 3. agents. In this way, moving from their definition, our account would correspond to the third level of the proposed taxonomy. From a certain standpoint, their definition applies at the second level of the taxonomy, to such an extent that we can use their definition of agency as a criterion for selecting systems to which our definition applies. Our account would then correspond to a subclass of their definition. However, their notion of normativity is stronger than ours, in this way designating a less minimalist account of agency with respect to ours.  

\

\noindent We now indicate three contexts in which such an account of agency may be fruitfully applied: 

\begin{enumerate}
\item \emph{Agency in the context of Quantum Mechanics (QM)}:  A possible way to put our conceptual strategy to the test is to offer a taxonomy of the various epistemic interpretations based on the underlying characterization of agency, developing further the analysis in (Barzegar \& Oriti, 2022). In this context, the point is to understand and characterize more precisely what is meant by the relational nature of quantum states. For instance, if one confines them to be dependent on a classical measurement context, no agency is involved in defining such context, and then Bohr’s interpretation is the one that fits best. If an agent is taken as an observer with a stronger form of agency (possessing personal experiences), then QBism represents the corresponding interpretative framework. If we weaken the status of the agent, naturalize it to identify an ‘observing system’ that can be a suitably complex physical system, accounting also for further elements of the context of interaction with the quantum system, and embrace a form of pragmatism, we get closer to interpretations \emph{à la} Healey or \emph{à la} Brukner-Zeilinger. Finally, if we broaden the notion of agency and ‘observer’ to such an extent of including every possible interacting physical system, then you get Rovelli's relational quantum mechanics. The latter would seem to be excluded by our minimalist definition of agency, if any agency is actually required for a relational view of quantum states. Our analysis of agency, and further refinements of the same, therefore, will correspond to an analysis of the differences between all these interpretations of quantum mechanics. 

\item \emph{Agency in the context of the debate about the nature of laws}: developing a more detailed account of agency might offer fruitful insights on the nature of laws. The idea would be to promote an account of laws in which the epistemic components are defining features of the laws themselves. Starting from the Better-Best System Account (BBSA) developed by Cohen and Callender (2009) \footnote{On this account, laws of nature are the axioms or theorems of the system achieving the best balance between simplicity and strength. However, it seems that simplicity and strength can only be evaluated based on epistemic features of the cognitive agents, and their interests. The BBSA explicitly acknowledges this fact. Laws are constructed by agents to be used by agents. Lawhood should be epistemically accessible. So, BBSA clearly makes laws dependent on agents and their epistemic features.}, the goal would be to deny the (tenability or usefulness of the) distinction between an objective reality and the content of a theory or set of theories. On this perspective, laws are produced in a perspectival manner, to such an extent that models and theories would not be connoted in terms of truthfulness or correspondence, rather in terms of similarity. Perception itself would be a kind of modelling, and not a direct representation of the world (in the sense of isomorphic correspondence). 
The agent has thus an irreducible role in constructing reality, including its \lq laws\rq. An agent can only interact in a specific way with the world, thus constructing a model of the latter from its own perspective. By taking stock of Putnam’s analysis on the notion of internal realism (Putnamn, 1976), it is possible to assert that accessing the world outside a model or a theory is not possible: every access is from within. But a precise epistemic account of laws, in this direction, requires a more precise account of agency, with particular attention to cognitive/epistemic agents as a subclass within our minimalist account.

\item \emph{Agency and Bayesian probabilities}: There are several issues to be considered in this context, the general question being what characterizes a {\it Bayesian agent}, i.e. an entity that uses Bayesian probabilities in its interactions with the environment. The first one is the relationship between Bayesian probabilities and agency. For instance, a system can be said to use Bayesian probabilities even in the limited sense of applying a spam filter: the latter is able to establish, based on relative frequency information, what corresponds to an incoming email and what is a spam. But this is not enough to classify the spam filter as a Bayesian agent. A spam filter uses Bayes' theorem to estimate priors via relative frequencies of the words in a given email based on its past occurrence in spam and non-spam emails. It calculates, based on Bayes' rule, but it has no degree of belief corresponding to the priors and posteriors. The issue of representing probabilities is different from the one of using probabilities. How can we translate this use of Bayesian probabilities into a role for agency? What are the elements in model-building that correspond to using Bayesian probabilities? The only way one can ascertain whether a system is using Bayesian calculus is by looking at its behaviour, but there is a difference between being able to represent probabilistic behaviour \emph{vs} using the Bayesian probabilities. What is the role of future-orientedness? We leave all this to future work. 
\end{enumerate}

\noindent These three lines of inquiry are deeply entrenched. First of all, one needs to spell out the relationship between our account of agency and the general discussion on the laws of nature. In the context of QM, one should explore whether and under which perspective Quantum Mechanical probabilities can or should be interpreted as Bayesian probabilities. If that is so, only those systems capable of using Bayesian probabilities, i.e. Bayesian agents, could apply Quantum Mechanical laws or play the role of the second system (the “observer”) with respect to which quantum states are defined in a relational fashion. 

\

\noindent In addition, there is a relation between the complexity of the agent and the kind of probabilities the agent can support or use. Let us sketch how this link can be understood, without any pretense of completeness or precision of details. We have defined agency as modeling activity. At the minimal level, modelling is any kind of structure or categorization of data received from the interaction with the environment. A higher level of complexity is found in terms of the existence of feedback loops in the internal information-processing activity of the agent. At this level, there can only be dispositions or propensities defined as objective, or context-dependent probabilities. At a higher level, the agent becomes a self-referential system whose successive interactions depend on its own activity. At this level of complexity, the agent becomes forward-looking, relying on its past history, where the patterns of past input-output are stored. 
A certain pattern is selected based on the weights the agent assigns to them which results in an output in the form of an action. Here one can talk about objective Bayesian probabilities which are mainly assigned according to the available evidence.

\

\noindent Furthermore, the agent can imagine or compose patterns which were never experienced before in its interaction with the environment and can take advantage of them to simulate or predict its future behavior. The agent becomes forward-looking and deals with counterfactual scenarios. Here, the agent moves to the level of subjective Bayesian probabilities because, in assigning the probabilities to different imagined or expected patterns, it mainly relies on subjective factors (though past evidence still plays a role). The more the agent could construct and use hypothetical patterns within its model, the more it can be viewed as “creative”.  Two are the reasons why it is here possible to define Bayesian probabilities as different from dispositions. First, the evidence does not fully determine the assigned  probabilities. Second, here the context is not just a generic system, rather it is a complex and special agent which deserves to be regarded as a subject assigning probabilities. It is indeed a subject in the sense that it accords probabilities to stored patterns based on its own, individual features.

\section{Comparison with other accounts of agency}

Let us now compare our account of agency and others available in the literature: 

\begin{enumerate}

\item \emph{Briegel}: According to Briegel (2012), an agent is an information processing system equipped with a memory that allows for \textit{projective simulation}. In his words, a memory is “any kind of organ, or physical device, that allows the agent to store and recall information about past experience” and it grants “the agent to relate its actions to its past” (Briegel, 2012, p. 2). At the minimal level of internal information processing, the actions of the system remain simple reflexes to environmental stimuli. All the behavioral outputs of the system are just immediate responses to the received data from interaction with the environment. At a second level, the information stored within the memory undergoes projective simulation. A memory can be modelized as a stochastic network of clips, where clips are the basic units of memory corresponding to very short episodes which are sequences of remembered percepts and actions. Triggered by a perceptual input, some specific clip in memory related to the input is excited. This excited clip will then, with a certain degree of probability, excite a neighboring clip, leading to a transition within the clip network. This process of generating excited clips continues until a certain feature is detected and thereby an excited clip couples out and translates into motor action. At a still higher level of internal information processing, there are not only transitions between existing clips, but the latter may themselves be \textit{created} and varied as part of the simulation process itself. This composition of new fictitious clips allows for simulation of plausible future experience and new patterns for future action. This allows the agent to detach itself from strict dependence on its surrounding environment by generating a space of possibilities for responding to environmental stimuli. Briegel’s notion of agency is very close to ours. More specifically, his notions of memory space and projective simulation correspond to our concept of modeling activity, thus we would argue that our definitions agree in making agency defined by “action based on internal modelling”. However, his definition is not as minimal and scalable as ours, because his notion of modelling is not as minimal as ours. Moreover, he invokes higher-order concepts like memory and experience which also distances his notion of agency from our naturalized account, placing it higher up in our scale of further classification of kinds of agency within our broader definition.

\item \emph{Hartle}: The concept of an information gathering and utilizing system (IGUS) was introduced by Hartle (2005) and has been further developed by Ismael (2011) and Callender (2017). An IGUS, as its name implies, is any system with the ability to gather information from its environment and then process this information to guide its future behavior. An IGUS has a memory register $P$, in which it stores the information received from the environment. This information is stored in $n+1$ memory registers with $P_0$ being the most recent data and $P_n$ the oldest ones. Each register is a snapshot of a different instant of time. When the system acquires new information, it is registered in $P_0$ and the data are conveyed to the memory registers, i.e, $P_n$ is erased and replaced by the information in $P_{n-1}$ and so on. In this way, the IGUS acquires a coarse-grained picture of its surrounding environment. The IGUS makes use of two kinds of information processing. Via a process called $U$, it builds a model of the world based on the information stored in its memory register. Using another process $C$, it takes the newest information stored in $P_0$ together with the model provided by $U$ to make predictions and act accordingly. The concept of an IGUS comes very close as well to our notion of an agent, and it is clearly based on the same conceptual and information-theoretic elements. However, the main difference arises from the the fact that while an IGUS only uses one model of the world, our definition of agency invokes a hierarchy of models, thus being more scalable and minimal. It follows that the concept of an IGUS can be regarded as identifying a further kind within our broader definition.

\item \emph{Moreno and Mossio}: The notion of agency developed by Moreno \& Mossio (2015) follows the tradition of self-maintenance accounts.\footnote{See, for instance, Bickhard \& Terveen (1995) and as a recent account Jones (2017).} On this perspective, the main focus lies within the context of biological autonomy, to such an extent that agency is primarily founded on the distinction between a system (the agent) and the environment. In addition, agents are connoted by their capacity to exert causal influence and operate via some normative principles. It is from these assumptions that Moreno \& Mossio ground their notion of agents as those systems which are able to maintain their own organization (also) through their influence on the surrounding environment, based on some notion of normativity. According to this account, while bacteria and higher organisms count as agents, viruses do not, in that the latter lack the organization which is required to bring about the proper interaction with the environment. Viruses do not display \lq\lq agential capacities\rq\rq, in the sense that “they exhibit such capacities only insofar as they are integrated into much more complex systems (typically cells) that are organised, in the specific sense that they realise a closure of constraints” (Moreno \& Mossio, 2015, p. 96). Their account is thus close to Barandiarian et al., but is definitely distant from ours. Indeed, as our minimalist notion of agency is explicitly scalable, we do not think it is necessary for a system to count as an agent that the distinction between the latter and the environment is clear and fixed once and for all. Rather, we claimed that this distinction can be context-dependent, thus defined in a perspectival manner. In addition, their notion of normativity is definitely more burdensome than ours. 

\item \emph{Rovelli}: In relational quantum mechanics (RQM) \footnote{Originally developed by Rovelli (1996), an introductory discussion is Laudisa \& Rovelli (2021).}, a quantum state ($\psi$) provides the information an observing system ($A$) gathers of an observed system ($S$). As this information is collected via the interaction between the two systems, the quantum state $\psi$ is relative to $A$. In this sense RQM is said to provide perspectival, system-dependent information, thus rebutting the idea of absolute, intrinsic properties of systems, as well as a notion of absolute and objective evolution through time.\footnote{Importantly, no hidden variables are posed, meaning that the quantum state is taken as a complete characterization of the physical system.} Quantum states of a system are only defined relationally, namely with respect to another system (called an observer or a reference-system) that is in relation with the former; in fact, this general relational, perspectival and context-dependent view on quantum states is common to a number of interpretations of quantum mechanics, collectively labelled \lq epistemic-pragmatist\rq or \lq neo-Copenhagen\rq. We refer to Brukner (2021), Pienaar (2021), as well as Di Biagio \& Rovelli (2021), for a discussion on the role of observers in RQM, and to Barzegar \& Oriti (2022) for a broader account of the general interpretative framework for QM, common to RQM and other relational or pragmatist perspectives. 
Importantly, in RQM the observing or reference-system can in principle be any other physical system -- where the latter, again, is not defined intrinsically, but always as a set of relations with respect to other systems. Moreover, as all systems are equivalent, it follows, as outlined by Van Fraassen (2010), that RQM provides a general expression for the information that any system can have of any other system. For what matters in the present discussion, the only way a notion of agency can be associated to the ‘observing system’ in RQM, not only it would make no reference at all to consciousness, but it would actually have to be minimized to include every physical system, no matter how elementary. This notion of agency has been in fact proposed by Rovelli (2020). Our definition of agency is similarly naturalized, and similarly based on an information-theoretic perspective, and in fact it would also suggest (or at least be consistent with) a relational understanding of quantum states, when applied as a basis for an interpretative framework for quantum mechanics. However, albeit minimalist, our definition of agency would not apply to any interacting physical system, and exclude several systems that would instead counts as agents according to Rovelli (2020). To the extent in which a notion of agency is considered useful or even needed, as an attribution of the ‘observing system’ in quantum mechanics, to specify the (relational, perspectival or context-dependent) nature of quantum states, then our definition would go in the direction of supporting epistemic-pragmatist interpretations other than RQM, such as Healey's or Brukner-Zeilinger, while not necessarily favouring the more radical direction of QBism. 

\item \emph{Steward}: Helen Steward is interested in defining a notion of agency that can be sufficiently loose to include animals. On her account, some criteria need to be fulfilled, namely that an agent: i. is able to move a whole, or some of its sub-components, as something we would call its \emph{body} (or parts of its body); ii. displays some form of \emph{subjectivity}; iii. exhibits at least some rudimentary form of \emph{intentionality}; iv. controls certain of the movements of its body (Steward, 2009, p. 226). This account of agency relies on a notion of intentional states as goal-orientedness, as well as on the idea that the agent is able to make choices. This account of agency, which is invoked by Steward to defend a libertarian metaphysics in the context of the free-will debate, is definitely stronger than our minimalist account and largely tangential to it.

\item \emph{Tononi et al.}: The starting point of Tononi and collaborators' analysis is that the reductionist paradigm proves untenable to account for autonomy and agency, in that the latter implies that there is a distinction between agents and environment, that agents are associated with macroscopic systems, and that agents are said to act upon the surrounding environment. Provided that a minimal notion of agency implies the existence of a system that “dynamically and informationally interacts with an environment” (Albantakis et al., 2020, p. 1), the subtler question is whether a subset of a larger system -- which is characterized by informational features -- can be qualified as an agent. It is for this reason that Tononi and collaborators invoke the so-called integrated information theory (IIT) paradigm: originally proposed as a theory of consciousness (Tononi, 2015), IIT provides a quantitative framework to specify the causal structure of discrete dynamical systems. Within this account, there is a quantity, called $\Phi$, that establishes the extent to which the causal constraints a certain system exerts on itself is irreducible to its underlying components.  IIT is defined from the intrinsic perspective of the system only if “this information is maximally irreducible \emph{both} in the past \emph{and} in the future” (Tononi, 2012, p. 301). 
It is argued that whenever $\Phi>0$, then the system is said to form a unitary whole, in the sense that “all of its subsets constrain and are being constrained by other subsets within the system above a background of external influences” (Albantakis et al., 2020, p. 2).\footnote{For further discussion on the topic, see, among others, Maturana and Varela, (1980), Marshall et al., (2017), Albantakis (2018).} Keeping this in the background, information is here defined in terms of alternatives, whereby a system can be qualified as an agent in case it is able to decide between a set of alternatives -- where this decision cannot be reduced to the causes of the underlying components. Moreover, Tononi relies on a minimal notion of consciousness, where minimality is granted by the fact that, for instance, the advanced notion does not imply self-awareness. Despite the minimality of the invoked notion of consciousness, we believe agency should be defined independently of a putative characterization of consciousness. Indeed, we claim that developing an account of agency in terms of modeling capacity makes it more naturalistic, scalable and, most importantly, minimal. However, the approach by Tononi and collaborators, which is based on information acquisition and processing, as well as the role they assign to the system's activity, is clearly consonant to ours. 

\item \emph{Winning}: Again, our account associates modeling capacity with agency. In this sense, it strongly resonates with Winning's (2020) analysis, according to which the notion of internal perspectivalism is preliminary to the possibility of defining an agent. His underlying assumptions are two. The first is a form of realism with respect to the existence of agents. The second is that agency is not a natural kind, in the sense that no fixed boundary between what is an agent and what is not can be ultimately defined. More specifically, he takes agents as a specific kind of autonomous control systems -- where an autonomous control system is one that displays a degree of influence on the surrounding environment which is above a certain threshold, a threshold that the system is able to recognize. To count as autonomous normative systems -- thus, as agents -- controllers should display not only a discrete output repertoire, but also a discrete input repertoire. Put it otherwise, they must be observers. And here is where Winning's account approaches ours. For a system displaying observational properties it must rely on some modeling activity. Indeed, there are many ways in which models can be related to autonomous control systems. In his PhD dissertation (Winning, 2019), Winning defines three ways controllers involve models. The first is the one drawn by Pattee (1995), according to which every controller mapping inputs for behavioral outputs defines an implicit model. The second is the adoption of models to retain information about the environment not presently tracked down, as in the case of forward models. The third is the use of models to account for hypothetical scenarios. On this perspective, an agent is a specific type of autonomous controller that is multidimensional, multivariate, composite, variable, and metamorphic (Winnig, 2019, pp. 125-126). Despite being close to our account, Winning's is still different, and definitely less minimalist than ours. Indeed, he takes agency as “control by means of preferences” (PhD thesis, p. 127), to such an extent that being an agent implies the possession of specific features geared towards preference-based behavior. In some more detail, an agent is a system that displays a perspective on the surrounding environment, carves it into a number of states of affairs and offers a perspective on what states of affairs are at stake and which ones are under its own control. The set of categorizations of such states of affairs, together with the agent's beliefs and preferences constitute its normative standard for how to control the surrounding environment. In this sense, Winning's account is closer to Sterenly's (2003), despite being more refined than the latter. It could be made closer also to ours only to the extent in which her notion of preferences and perspective can be scaled down\rq to a more minimalist and naturalised level. 
 
\end{enumerate}

\section{Objections}

In this section we consider a number of objections that can be raised against our account. 

\

\noindent First of all, one might be concerned that our account relies too heavily on the representationalist paradigm, according to which representation plays a central role in the way agents acquire knowledge or information about the environment. In other words, one might interpret the notion of modeling as constructing a copy of the surrounding environment from the inside. Then the agent uses this internal model to interact with and navigate such an environment. However, according to the embodied cognition paradigm (see, among others, Mahon, 2015; Shapiro, 2019), this is not the way agents interact with and acquire knowledge or information about the world. According to the embodied paradigm, in contrast to the representationalist paradigm, the priors in perceptual anticipation are not beliefs and assumptions.  Rather, they are embodied skills and patterns of action-readiness. The agent does not have a model of the world. Rather, it is a model of the world. 

\noindent The controversy regarding the representationalist \textit{vs} embodied paradigms of cognition is a debate concerning whether one should regard the brain or the body as the locus of cognitive activities. Here, we are just interested in portraying a minimalist notion of agency, thus remaining neutral with respect to such a debate. It can be the body that acts as the modeling element, in the sense that it displays a form of covariance between its activities and the environmental stimuli. Alternatively, it can be the brain that acts as the modeling element, in the sense that there is a covariance between its neuronal activities and the external stimuli. And importantly, in both cases the modeling element can, under appropriate conditions, serve the purpose of constructing hypothetical scenarios geared towards successive interaction with the environment. Furthermore, we are concerned with a minimal notion of agency and representation not loaded with cognitive and conscious activities. Of course, at the higher levels of modeling activity, one can introduce considerations on cognition or consciousness according to our relational notion of hierarchy of models. Still, at the minimal level of data models, our notion of agency has nothing to do with a strong notion of representation, as used within cognitive science.

\

\noindent A second objection that can be raised, which directly follows from the first one, is that our notion is too minimal, and in this way is not able to capture the essential character typically associated to agency. As already discussed in the introduction, the traditional notion of agency is connected to the idea of a system endowed with the capacity to act -- where the latter is constructed in terms of intentional states exerted by the agent's mental faculties. All in all, it looks as if we have betrayed the traditional apparatus concerning agency. In this sense, accounts such as Tononi's and recently Ried et al.'s (2019) come more equipped to address such traditional requirements. For instance, Ried et al. (2019)'s definition of agents is “entities that interact with their environment via explicitly modelled perceptions and actions, endowed with an internal mechanism for deciding how to respond, and capable of adapting those responses based on an individual history of interactions and feedback” (p. 2). Two comments can be added. The first one is that we think disentangling agency from notions such as intentionality, perception and mental states is necessary if we aim at naturalizing it, at least to some extent. The second one is that, as stressed time and again throughout this paper, our notion of agency is scalable, and can thus be structured in a hierarchy of levels, the higher ones of which may introduce some of the other features that are traditionally associated to agency. Finally, our goal was indeed to achieve a minimalist account of agency that was at the same time not so minimal as to be trivial or to end up regarding any possible physical system as an agent. We think this goal was achieved, and we regard our notion of agency as indeed very minimal (leaving out features like the ones discussed above) but not too minimal being not obviously satisfied in concrete cases.

\

\noindent A third objection is the opposite of the previous one, and would arise, for example, in approaches aiming at erasing any demarcation among physical systems, for example in the context of relational approaches to quantum physics (and maybe physics in general). The notion of agency that could be attributed to the observing or reference system in RQM, as discussed, would have to be weaker than ours, since it has to fit any interacting physical system. 
However, within this context, we would argue that the resulting notion of agency is so much thin, minimalist, and objectified that one may wonder what is the reason to introduce it in the first place, namely why to distinguish it from simple interaction. 
Regardless of the respective merits of RQM {\it vis-à-vis} other interpretations of quantum mechanics, we would maintain that such extremely weak notion of agency can hardly serve the purpose of providing fruitful insights in the contexts we outlined -- such as the debate on the nature of physical laws and bayesianism. In fact, we would argue that it does not even help in the interpretation of quantum mechanics; being so weak, it does not help to elucidate any aspect of quantum mechanics that RQM does not elucidate already without it; therefore, it cannot be the basis for preferring RQM over its alternatives.

\section{Concluding remarks}
In this paper, we have proposed a minimalist, scalable and naturalized account of agency. To this end, agency has been related to the capacity of a certain system to display modeling activity -- where minimality is guaranteed by the fact that even a simple classification of the information arising from the environment can be taken as a kind of model-building (of data models, more precisely). We have explored some possible refinements of our definition, so to assure its scalability. We have discussed how our notion compares with other accounts of agency in the literature, emphasizing the similarities with those that also are based on information acquisition and processing. Finally, we have pointed to some contexts -- such as the interpretation of quantum mechanics, the debate on the nature of physical laws and bayesianism in physics, epistemology and philosophy of science (see Bovens \& Hartmann, 2003 and Hartmann \& Sprenger, 2019 for general overviews)-- in which our account may provide fruitful insights. We leave the application of our account of agency to these contexts for future work.

\section*{Acknowledgements}
This research has been funded by the FQxI Grant FQXi-RFP-IPW-1908. DO acknowledges financial support also from DFG research grants OR432/3-1 and OR432/4-1. EM thanks MCMP and LMU for hospitality during the research work for this project.


\begin{thebibliography}{9}


\bibitem{Albantakis: 2018}
Albantakis, L. (2018). A Tale of Two Animats: What Does It Take to Have Goals? Springer, Cham: 5–15.

\bibitem{Albantakis:2020}
Albantakis, L., Massari, F., Beheler-Amass, M., Tononi, G. (2020). A macro agent and its action. arXiv:2004.00058 [q-bio.NC], https://doi.org/10.48550/arXiv.2004.00058.

\bibitem{Anscombe:1957}
Anscombe, G.E.M. (1957). \emph{Intention}, Oxford: Basil Blackwell.

\bibitem{Barandiaran:2009}
Barandiaran, X.E., Di Paolo, E., Rohde, M. (2009). Defining Agency: Individuality, Normativity, Asymmetry, and Spatio-Temporality in Action. \emph{Adaptive Behavior}, 17(5): 367–386.

\bibitem{B+O:2022}
Barzegar, A., Oriti, D. (2022). Epistemic-Pragmatist Interpretations of Quantum Mechanics: A Comparative Assessment. arXiv:2210.13620 [quant-ph]. 

\bibitem{B+T:1995}
Bickhard, M. H., \& Terveen, L. (1995). \emph{Foundational Issues in Artificial Intelligence and Cognitive Science: Impasse and Solution}. Amsterdam: North-Holland.

\bibitem{B+I+D:2016}
Biehl, M., Ikegami, T., Polani, D. (2016). Towards information based spatiotemporal patterns as a foundation for agent representation in dynamical systems. 	arXiv:1605.05676 [cs.AI].

\bibitem{Bokulich}
Bokulich, A. (2011). How scientific models can explain. \emph{Synthese}, 180(1): 33–45.

\bibitem{Bovens}
Bovens, L., Hartmann, S. (2003) \emph{Bayesian Epistemology}, Oxford: Oxford University Press.

\bibitem{Briegel:20}
Briegel, H. J. (2012). On creative machines and the physical origins of freedom. \emph{Scientific Reports} 2: 522.  

\bibitem{Brukner:2021}
Brukner, C. (2021). Qubits are not observers -- a no-go theorem. arXiv:2107.03513 (2021).

\bibitem{Cartwright:1983}
Cartwright, N. (1983). \emph{How the Laws of Physics Lie}. Oxford: Oxford University Press.

\bibitem{C+C:2009}
Cohen, J., Callender, C. (2009). A better best system account of lawhood. \emph{Philosical Studies}145:1–34.

\bibitem{Callender:2017}
Callender, C. (2017). \emph{What makes time special?}. Oxford University Press.

\bibitem{Curiel:2020}
Curiel, E. (2020). Schematizing the Observer and the Epistemic Content of Theories. arXiv:1903.02182 [physics.hist-ph]

\bibitem{Davidson:1963}
Davidson, D. (1963). Actions, Reasons, and Causes, reprinted in Davidson 1980: 3-20.

\bibitem{Davidson:1980}
Davidson, D. (1980). \emph{Essays on Actions and Events}, Oxford: Clarendon Press.

\bibitem{D+R: 2021}
Di Biagio, A., Rovelli, C. (2021). Relational Quantum Mechanics is About Facts, Not States: A
Reply to Pienaar and Brukner. \emph{Foundations of Physics} 52 (3), 62, arXiv: 2110.03610 [quant-ph].

\bibitem{F+H:2020}
Frigg, R., Hartmann, S. "Models in Science", The Stanford Encyclopedia of Philosophy (Spring 2020 Edition), Edward N. Zalta (ed.), URL = <https://plato.stanford.edu/archives/spr2020/entries/models-science/>.

\bibitem{F+N:2020}
Frigg, R., Nguyen, J. (2020). \emph{Modelling Nature: An Opinionated Introduction to Scientific Representation}. Synthese Library 427. 

\bibitem{F+N:2021}
Frigg, R., Nguyen, J. "Scientific Representation", The Stanford Encyclopedia of Philosophy (Winter 2021 Edition), Edward N. Zalta (ed.), URL = <https://plato.stanford.edu/archives/win2021/entries/scientific-representation/>.

\bibitem{Giere:1988}
Giere, R. N. (1988). \emph{Explaining Science: A Cognitive Approach}. Chicago: University of Chicago Press.

\bibitem{Giere:2010}
Giere, R. N. (2010). An agent-based conception of models and scientific representation. \emph{Synthese} 172:269–281

\bibitem{Hacking:2002}
Hacking, I. (2002). \emph{Historical Ontology}. Cambridge: Cambridge University Press.

\bibitem{Hacking:2009}
Hacking, I. (2009). \emph{Scientific Reason}. Taipei: National Taiwan University Press. 

\bibitem{Hartle:2005}
Hartle, J. B. (2005). The physics of now. \emph{American Journal of Physics}, 73(2):101-109.

\bibitem{Hartmann:2019}
Hartmann, S., Sprenger, J. (2019). \emph{Bayesian Philosophy of Science}, Oxford: Oxford University Press. 

\bibitem{H+M:1978}
Hegner, S. J., Maulucci, R. A. (1978). Set-theoretic foundations of data-structure representation. \emph{Inform. Systems} 3: 193-201. 

\bibitem{H+A+T:2013}
Hoel, E. P., Albantakis, L., Tononi, G. (2013). Quantifying causal emergence shows that macro can beat micro. PNAS 110 (49) 19790-19795 https://doi.org/10.1073/pnas.131492211.

\bibitem{H+M:2014}
Hutto, D., Myin, E. (2014). Neural Representations Not Needed: No More Pleas, Please. \emph{Phenomenology and the Cognitive Sciences}, 13(2): 241–256.

\bibitem{Isaac:2013}
Isaac, A. M. C. (2013), Modeling without Representation, \emph{Synthese}, 190(16): 3611–3623.

\bibitem{Ismael}
Ismael, J. (2011). “Decision and the open future”. In Bardon, A. (Ed.), \emph{The future of the Philosophy of Time}, London, Routledge, pp. 149-168.

\bibitem{Jones:2017}
Jones, D. M. (2017). The Biological Foundations of Action. London: Routledge.

\bibitem{King:2016}
King, M. (2016). On structural accounts of model-explanations, \emph{Synthese} 193:2761–2778.

\bibitem{L+R:2021}
Laudisa, F., Rovelli, C. (2021). "Relational Quantum Mechanics", The Stanford Encyclopedia
of Philosophy (Spring 2021 Edition), https://plato.stanford.edu/archives/spr2021/entries/qm-relational/.

\bibitem{Leonelli:2016}
Leonelli, S. (2016). \emph{Data-centric biology: A philosophical study}. Chicago: Chicago University Press.

\bibitem{Leonelli:2019}
Leonelli, S. (2019). What distinguishes data from models? \emph{European Journal for Philosophy of Science} 9:22. https://doi.org/10.1007/s13194-018-0246-0.

\bibitem{Levy:2015}
Levy, A. (2015). Modeling without Models, \emph{Philosophical Studies}, 172(3): 781–798. 

\bibitem{Mahon:2015}
Mahon, B. Z., 2015. What Is Embodied about Cognition?. \emph{Language, Cognition and Neuroscience}, 30(4): 420–29. 

\bibitem{M+W+T*A:2017}
Marshall, W., Kim, H., Walker, S.I., Tononi, G., Albantakis, L. (2017) How causal analysis can reveal
autonomy in models of biological systems. \emph{Philosophical Transactions of the Royal Society A. Mathematical, Physical and Engineering Sciences} 375:20160358.

\bibitem{Massimi:2018}
Massimi, M. (2018). Perspectival Modeling, \emph{Philosophy of Science}, 85(3): 335–359. 

\bibitem{M+V:1980}
Maturana, H. R., Varela, F. J. (1980). \emph{Autopoiesis and Cognition. The Realization of the Living.} 
Boston studies in the philosophy of science (42). 

\bibitem{Morrison:1999}
Morrison, M. (1999), “Models as Autonomous Agents”. In Morgan and Morrison (eds.) Morgan, \emph{Models as Mediators: Perspectives on Natural and Social Science}, Cambridge, Cambridge University Press: 38–65. 

\bibitem{Pattee:1995}
Pattee, H. H. (1996). The Problem of Observables in Models of Biological Organizations. Reprinted in H. H. Pattee \& J. Rączaszek-Leonardi (Eds.), Laws, Language and Life:
Howard Pattee’s Classic Papers on the Physics of Symbols with Contemporary Commentary (pp. 245–259). Dordrecht: Springer, 2012.

\bibitem{Pienaar:2021}
Pienaar, J. L. (2021). A quintet of quandaries: five no-go theorems for Relational Quantum Mechanics. arXiv:2107.00670 [quant-ph]

\bibitem{Putnam:1976}
Putnam, H. (1976). What Is "Realism"? \emph{Proceedings of the Aristotelian Society} 76:177-194. 
 
\bibitem{Riedetal:2019}
Ried, K., M\"uller, T., Briegel, HJ (2019) Modelling collective motion based on the principle of agency: General framework and the case of marching locusts. \emph{PLoS ONE} 14(2): e0212044. https://doi.org/10.1371/journal.pone.0212044.

\bibitem{Rovelli:1996}
Rovelli, C. (1996). Relational quantum mechanics. \emph{International Journal of Theoretical Physics} 35: 1637–1678. 

\bibitem{Rovelli:2016}
Rovelli, C. (2016). Meaning = Information + Evolution. arXiv:1611.02420v1 [physics.hist-ph]. 

\bibitem{Rovelli:2018}
Rovelli, C. (2018). Space is blue and birds fly through it. \emph{Philosophical Transactions of the Royal Society A: Mathematical, Physical and Engineering Sciences}, 376(2123), 20170312.

\bibitem{Rovelli:2020}
Rovelli, C. (2020). Agency in Physics. arXiv:2007.05300 [physics.hist-ph]. 

\bibitem{Schlosser:2019}
Schlosser, M. (2019). Agency. \emph{The Stanford Encyclopedia of Philosophy (Winter 2019 Edition)}, Edward N. Zalta (ed.), URL = <https://plato.stanford.edu/archives/win2019/entries/agency/>.

\bibitem{Shapiro:2019}
Shapiro, L. (2019). \emph{Embodied Cognition}, Second Edition, London; New York: Routledge.

\bibitem{S+C:2011}
Silberstein, M.,  Chemero, A. (2011). Dynamics, Agency and Intentional Action, \emph{Humana Mente}, 15: 1–19.

\bibitem{Sterelny:2003}
Sterelny, K. (2003). \emph{Thought in a Hostile World: The Evolution of Human Cognition}. Malden, MA: Blackwell.

\bibitem{Steward:2009}
Steward, H. (2009). Animal Agency. Inquiry, 52(3), 217–231.

\bibitem{Tononi:2012}
Tononi, G. (2012). Integrated information theory of consciousness: an updated account. \emph{Archives Italiennes de Biologie} 150: 290-326. 

\bibitem{Tononi:2015}
Tononi, G. (2015) Integrated information theory. \emph{Scholarpedia} 10:4164.

\bibitem{V:2021}
Verreault-Julien, P. (2021). Factive inferentialism and the puzzle of model-based explanation. \emph{Synthese} 199: 10039–10057.

\bibitem{Weisberg:2012}
Weisberg, M. (2013).Getting Serious about Similarity. \emph{Philosophy of Science}, 79(5): 785-794. 

\bibitem{W:2019}
Winning, J. (2019). \emph{The Mechanistic and Normative Structure of Agency}. PhD Thesis dissertation. 

\bibitem{W:2020}
Winning, J. (2020). Internal Perspectivalism: The Solution to Generality Problems about Proper Function and Natural Norms. \emph{Biology \& Philosophy}, 35(33), 1–22. 

\bibitem{Y:2007}
Yannakoudakis, E. J. (2007). A Set-Theoretic Data Model For Evolving Database Environments. \emph{Proceedings of the 2007 International Conference on Information \& Knowledge Engineering}.

\bibitem{Zhao:2019}
Zhao,H. (2019). Category theory for (big) data modeling and model’s transformation. \emph{Databases [cs.DB]}. Université de Haute Alsace - Mulhouse, 2019. English. ffNNT : 2019MULH2946ff. fftel-03704110f 

\end{thebibliography}
\end{document}